\documentclass[twocolumn,prl,aps]{revtex4}
\usepackage{amsmath}
\usepackage{graphicx}
\usepackage{sistyle}

\begin{document}

\title{Generation of tunable wavelength coherent states and heralded single photons for quantum optics applications.}

\author{N.~Bruno, A.~Martin, and R.~T.~Thew}
\affiliation{Group of Applied Physics, University of Geneva, Switzerland
\\}

\begin{abstract}
Quantum optics experiments frequently involve interfering single photons and coherent states. In the case of multi-photon experiments this requires that all photons are frequency degenerate. We report a simple and practical approach to generate coherent states that can be readily tuned to any wavelength required, for example by non-degenerate photon pair creation. We demonstrate this by performing a two-photon (Hong-Ou-Mandel) interference experiment between a coherent state and a pure heralded single photon source. No spectral filtering is required on either source, the coherent state constrained by the pump and seed lasers and the heralded photon exploits non-local filtering. We expect that such an approach can find a wide range of applications in photonic based quantum information science.
\end{abstract}

\maketitle

\section{Introduction}

The interplay between coherent states and single photons has a rich history. Some of the first interference experiments~\cite{Tapster1998,Rarity2005} and Bell tests~\cite{Pittman2003a} between independent sources and the characterization of a single photon states using homodyne detection \cite{Babichev2004} exploited these different photonic resources. Recently, the combination of single photons and homodyne detection~\cite{Lvovsky2013}, or displacement operations~\cite{Bruno2013}, has opened up new avenues for testing fundamental questions about entanglement and macroscopic systems, as well as witnessing single photon entanglement~\cite{Morin2013}. Coherent states have also been shown to be a useful resource for characterizing the purity of photon states~\cite{Cassemiro2010} or generating entangled coherent states and quantum metrology~\cite{Joo2011}. The combination of single photon systems and coherent states also opens up possibilities for hybrid systems involving continuous variable and discrete detection schemes~\cite{Sangouard2010}.

Typically, experiments of this type involve using a pump laser to generate the coherent states that are then interfered with photons pairs, for example from spontaneous parametric downconversion (SPDC) that is pumped by the frequency doubled laser. This approach constrains all the photons to be degenerate in frequency. In the case of multiple independent sources it is also important that the photons are in a pure single mode~\cite{Grice2001,Osorio2013}. In the context of an increasing range of applications, an interesting task is to develop schemes to efficiently generate perfectly indistinguishable coherent states and single photons with low loss, low noise and, importantly, with wavelengths tunable over a wide range. 

 In the following we describe how we can efficiently generate pure heralded single photon (HSP) states and interfere this with a wavelength-tunable coherent state, without spectral filtering. We show the indistinguishability of these two sources by performing a two photon interference experiment~\cite{Hong1987} between the two independent sources.

\section{Heralded Single Photon source}

There has been enormous progress in developing heralded single photon sources and engineering their characteristics to suit myriad applications~\cite{Grice2001,Mosley2008b,Clark2011,Pomarico2012,Spring2013}. Typically we are interested in photonic sources for quantum communication, therefore we need to ensure that the photons are pure, indistinguishable, narrowband and at telecom wavelengths, which has proven to be a more demanding challenge. The most straightforward way to herald pure single photons is use nonlocal filtering to prepare the heralded photon in a pure state~\cite{URen05}.

To realize the heralded single photon source we use pairs of photons produced via SPDC.  The Hamiltonian for the process has the form:
\begin{equation}
H = \zeta \int {d\rm \omega_s}{d \rm \omega_i }S{\rm (\omega_s, \omega_i) } a_{\rm s}^{\dagger}({\rm \omega_s}) a_{\rm i}^{\dagger}({\rm \omega_i})+ {\rm h.c.}
\end{equation}
where $S{\rm (\omega_s, \omega_i) }$ is the Joint Spectral Amplitude (JSA) of the photons, as a function of the signal (s) and idler (i) frequencies ($\rm \omega_{s,i}$). The process is governed by the conservation of energy and momentum between final and initial configuration:
\begin{equation}
\rm \omega_p = \omega_s + \omega_i \, ;
\end{equation}
\begin{equation}
k_{\rm s}({\rm \omega_s}) +  k_{\rm i}({\rm \omega_i})- k_{\rm p}(\omega_{\rm s}+\omega_{\rm i})- \frac{2 \pi}{\Lambda} = 0
\label{eq_phasematching}
\end{equation}
where $\Lambda$ is the poling period of the crystal. Periodical poling is a technique based on the inversion of the crystal polarization with a period $\Lambda$ that can be tuned and optimized in order to achieve the desired phase matching condition (quasi-phase matching), with a greater efficiency than typical phase-matched materials.

\begin{figure}[h!]
\centering
\includegraphics[width=0.8\columnwidth]{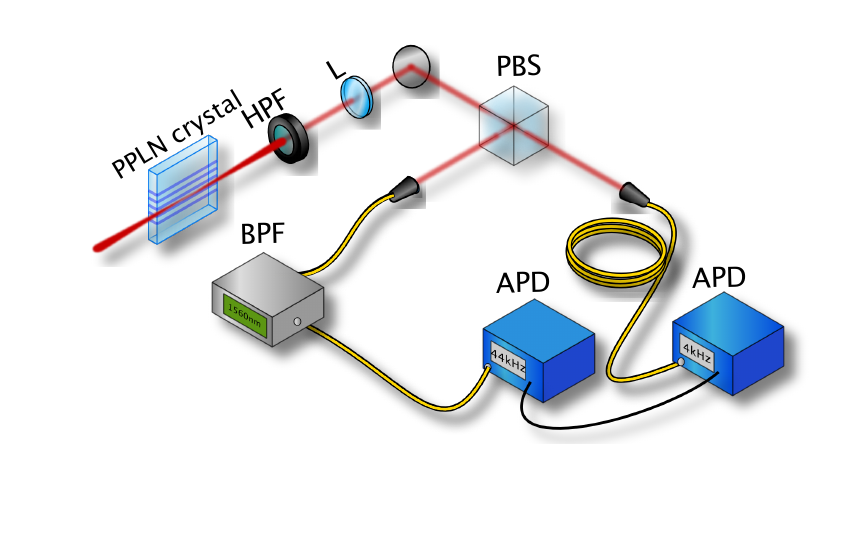}
\caption{Setup for the heralded single photon source based on a type II nonlinear crystal. The photon pairs generated by the crystal are deterministically separated by a polarizing beamsplitter (PBS) and coupled into a single mode fiber. A high-pass filter (HPF) and a band-pass filter (BPF) are employed, respectively, to remove the pump and to filter the herald photons upon detection by an avalanche photodiode (APD), thus projecting the second photon into a pure state.}
\label{fig_hsps}
\end{figure}

In our experiment, SPDC is achieved in a periodically poled Lithium Niobate (PPLN) bulk crystal, with type-II quasi-phase matching. To generate photon pairs at telecom wavelengths, the crystal is pumped at 780\,nm by a pulsed ($\Delta t =  2$\,ps) laser with a repetition rate of 76\,MHz. The two photons are generated with orthogonal polarization, thus they can be deterministically separated on a polarizing beam splitter (PBS). \figurename{~\ref{fig_hsps}} shows the setup for the Heralded Single Photon (HSP) source.

To observe two photon interference effects~\cite{Hong1987} with a good visibility we need to maximize the spectral and temporal overlap of the two photons. If the two photons are coming from different sources, this condition imposes the need for a single spectral and temporal mode, \textit{i.e.} to produce separable photons~\cite{Mosley2008b}. Narrow spectral filtering is the most common solution, but is not suitable for configuration where low loss is required~\cite{Aboussouan2010}.  We take advantage of the fact that selecting a single spectral mode on the heralding (idler) photon with a sufficiently narrowband filter, heralds the presence of a signal photon in the correlated mode. Hence, the signal photon is heralded in a pure state~\cite{URen05}, without any need for additional filtering, thus minimizing loss. The final efficiency of this heralding then only depends of the coupling efficiency~\cite{Guerreiro2013}.

The single- or multi-mode nature of the photon's state can be determined by measuring the second order autocorrelation function ($g^{(2)}(\tau)$) in a Hanbury Brown and Twiss like experiment~\cite{BROWN1956}. This measurement allows one to estimate the photon number statistics. For a single mode source the photon number distribution is thermal, and for a large number of independent modes with thermal statistics the photon number becomes Poissonian. The value of $g^{(2)}(0)$ is directly related to the purity (and, consequently, to the number of modes) from the relation $g^{(2)}(0)= 1+\mathcal{P} = 1+\frac{1}{K}$, where $\mathcal{P}$ and $K$ represent the purity and the number of Schmidt modes~
\footnote{ The Schmidt number is defined $ K = 1/ \sum_n \vert c_n \vert^4$ where $c_n$ is the coefficient of the Schmidt decomposition of the joint spectral amplitude of the pairs: $f(\omega_{\rm s},\omega_{\rm i})=\sum_n \Psi(\omega_{\rm i})\varPsi(\omega_{\rm s})$ (see Ref.~\cite{Mosley2008a} for more detail).}
, respectively. For a pure photon we expect a $g^{(2)}$ close to 2, which corresponds to thermal statistics. For a large number of modes, $K\longrightarrow  \infty$, we find that  $g^{(2)}(0)\longrightarrow 1$, which is the signature of a Poissonian distribution~\cite{Tapster1998,Sekatski2012}.

\begin{figure}[h]
\centering
\includegraphics[width=0.9\columnwidth]{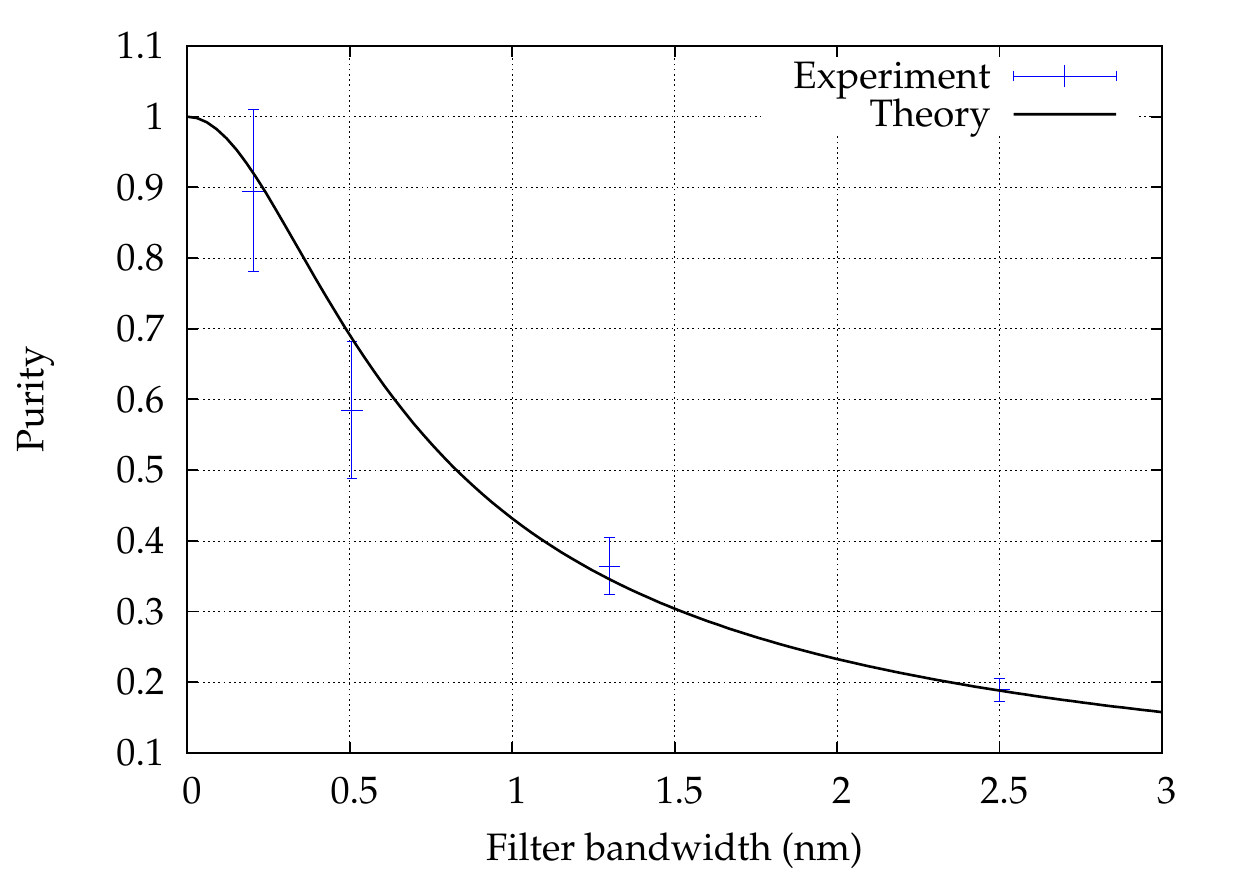}
\caption{\label{fig_g2} Purity of the signal photon as a function of the bandwidth of the filter placed on the idler photon arm. For a bandwidth of 0.2\,nm we obtain a purity of $0.9 \pm 0.1$.}
\end{figure}

 \figurename{~\ref{fig_g2}} shows a measurement of the purity $\mathcal{P}$ of the idler photon as a function of the filter bandwidth. The spectral correlation can be quantified evaluating the average number of Schmidt modes, which is found to be $K = 5 \pm 1$ without filtering. The spectral correlation is due a combination between the properties of the material and the bandwidth of the pump laser. 
Choosing a 0.2 nm filter for the idler photon allows one to achieve a purity of $0.9 \pm 0.1$ for both photons of the pair. The bandwidth of such heralded single photon is then limited by the pump bandwidth, due to energy conservation, and it is found to be equal to $79\pm1$ GHz.

\section{Tunable Coherent State source}

The purpose of this article is to demonstrate a tunable coherent state source that can be used to in conjunction with heralded single photons. This kind of interference has historically been realized doubling the frequency of a laser to  subsequently down-convert it and using the same laser as a local oscillator~\cite{Rarity2005,Pittman2003a,Pittman2005,Lvovsky2013}. Femtosecond pulsed lasers allow one to have a broad spectrum, such that filtering both heralded single photon and coherent state with similar band-pass filters is possible, although lossy. Indeed, to efficiently generate coherent states with the same temporal and spectral characteristics as the heralded photon is difficult. The solution we propose here consists in the use of a coherent state produced by means of difference frequency generation (DFG) in a nonlinear crystal, as in~\cite{Chou1998}. 
 
DFG takes place in materials with a $\chi^{(2)}$ nonlinearity and it is a stimulated process that happens in the presence of two strong coherent states, namely a pump laser and a seed laser. This is in contrast to SPDC, which is a spontaneous process that requires only one strong pump and is stimulated by the vacuum field. In DFG, energy and momentum conservation (Eq.~\ref{eq_phasematching}) also need to be satisfied:
\begin{equation}
\hbar \omega_p + \hbar  \omega_i = \hbar  \omega_s + 2 \hbar \omega_i \, ,
\end{equation}
where the frequency of the idler photon is given by the difference $\omega_p -\omega_s = \omega_i$.

\begin{figure}[h]
\centering
\includegraphics[width=0.8\columnwidth]{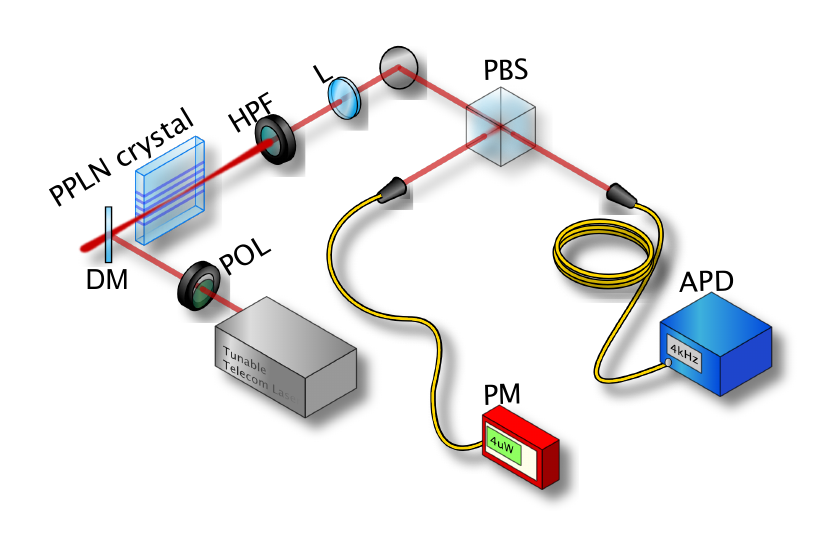}
\caption{Experimental setup of the coherent state source. A type-II nonlinear crystal is pumped by a  780\,nm pulsed laser and seeded by a CW laser at 1563.5\,nm, which allows one to stimulate the nonlinear interaction. As a result, we obtain a coherent state at 1556.5\,nm with a bandwidth of $80\pm1$\,GHz, given by the pump laser.}
\label{fig_coherent}
\end{figure}

The setup is shown in \figurename{~\ref{fig_coherent}}: the crystal we use is similar to the one used for the HSP, and the pump light is coming from the same laser. The seed for the nonlinear process is a laser at telecom wavelengths (1563.5\,nm). Since the latter is a monochromatic laser, the bandwidth of the coherent state resulting from the DFG process is fixed by the bandwidth of the pulsed pump.

\begin{figure}[h]
\centering
\includegraphics[width=0.9\columnwidth]{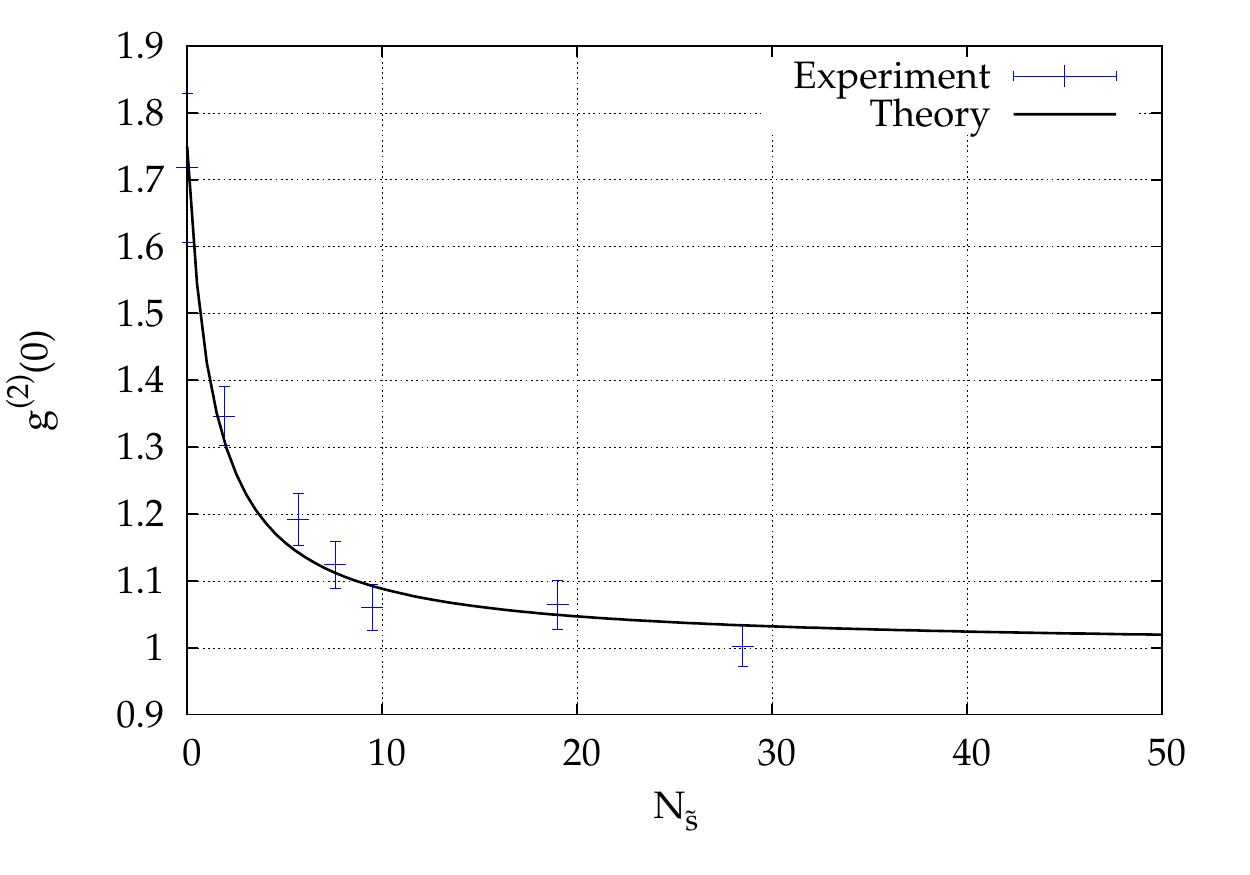}
\caption{$g^{(2)}(0)$ of the state at the output of the DFG as a function of the number of photon in the "seed" field for 10\,mW of pump. The solid curve represents the equation (\ref{eq_g2coherent}).
\label{fig_g2cs}}
\end{figure}

The coherent state is characterized in the same way as the HSP source, by measuring the second order autocorrelation function $g^{2}(0)$. For a coherent state $g^{2}(0)=1$. In the case of DFG, where the process is stimulated, this function becomes:
\begin{equation}
g^{2}(0) =  1+ \frac{\rm N_{SP}}{\rm K N_{SP}+N_{ST}} ,
\end{equation}
where $\rm N_{SP}$ and $\rm N_{ST}$ represent the number of photon per mode emitted by spontaneous emission and the number of photons emitted by stimulated emission, respectively. Moreover, in this configuration, where just one mode is stimulated $\rm N_{ST}$ is given by $\rm N_{ST} = N_{\tilde{s}} N_{SP}$ (see Ref.~\cite{Polyakov2009,Sanguinetti2012} for more detail), with $\rm N_{\tilde{s}}$ corresponding to the number of photons in the seed field. So, we obtain: 
\begin{equation}\label{eq_g2coherent}
g^{2}(0) =  1+ \frac{1}{\rm N_{\tilde{s}}+K}.
\end{equation}
In \figurename{~\ref{fig_g2cs}}, we see that as the input intensity of the seed field increases, the output idler state is coherent and the spontaneus emission is negligible. $N_{\tilde{s}}$ is chosen to be greater than 30 and an attenuator is placed at the output of the crystal to produce a coherent state $|\alpha\rangle$ with $\vert \alpha \vert^2~\ll~1$.

\section{Hong-Ou-Mandel interference}

\begin{figure}[h]
\centering
\includegraphics[width=1\columnwidth]{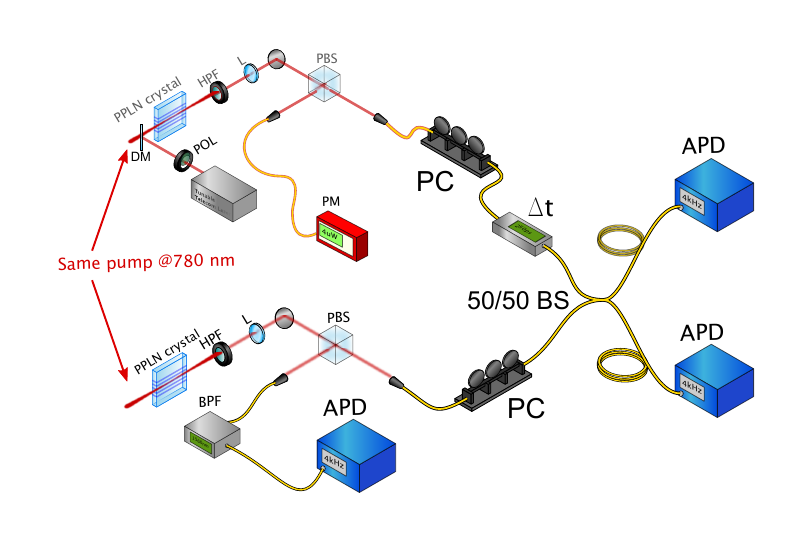}
\caption{Experimental setup  for the two photon interference between the HSP and the coherent state generate by SPDC and DFG processes, respectively. 
\label{fig_setupdip}}
\end{figure}

To prove the indistinguishability between the HSP and the coherent state generated by DFG, a two photon interference experiment is performed~\cite{Hong1987,Pittman2003a,Rarity2005,Pittman2005}. As presented in \figurename{~\ref{fig_setupdip}}, the two states are mixed on a 50/50 fiber coupler. To avoid polarization and temporal discernibility that could reduce the interference visibility, two Lef\`evre fiber polarization controllers and an adjustable optical delay line are employed, respectively. Note that the two interfering states never pass through a filter. The spectral overlap is only ensured by the filter placed on the correlated heralding photons, and the energy conservation for the coherent state. The absence of filters allows one to achieve a heralding coupling efficiency of 50\%. Losses are mainly due to fiber coupling at the output of the PPLN crystal. 

Furthermore, the visibility is also limited by the statistics of the sources. Two photons coming at the same time from one source can give a coincidence which reduces the dip visibility. Assuming that the probability of having three or four photons is negligible, the interference visibility is given by:
\begin{equation}\label{eq_maxvis}
\rm V_{max} = \frac{(1-\mathcal{R}) P_{1,a}P_{1,b} + P_{0,a}P_{2,b} + P_{2,a}P_{0,b}}{P_{1,a}P_{1,b} + P_{0,a}P_{2,b} + P_{2,a}P_{0,b}},
\end{equation}  
where $\rm P_{n,j}$ represents the probability to have n photon in the input arm $a$ or $b$ and $\mathcal{R}$ quantifies the indiscernibility between the two photons over all the observables. For a coherent state $|\alpha\rangle$ the probability is given by:
\begin{equation}
\rm P_{n,\alpha}= e^{-|\alpha|^2}\dfrac{|\alpha|^{2n}}{n!},
\end{equation}
and for the HSP by:
\begin{equation}
\left\{
\begin{array}{l}
\rm P_{0,HSP} = 1-(P_{1,HSP}+P_{2,HSP})\\
\rm P_{1,HSP}=t + 4 (t-1)t P_1\\
\rm P_{2,HSP}=2 t^2 P_1,
\end{array}
\right.
\end{equation}
where $\rm P_1$ and t are the probabilities to emit a pair and the photon's transmission through the system, respectively. These equations are valid if we assume a low detection efficiency for the heralding photons.

\begin{figure}[h]
\centering
\includegraphics[width=0.9\columnwidth]{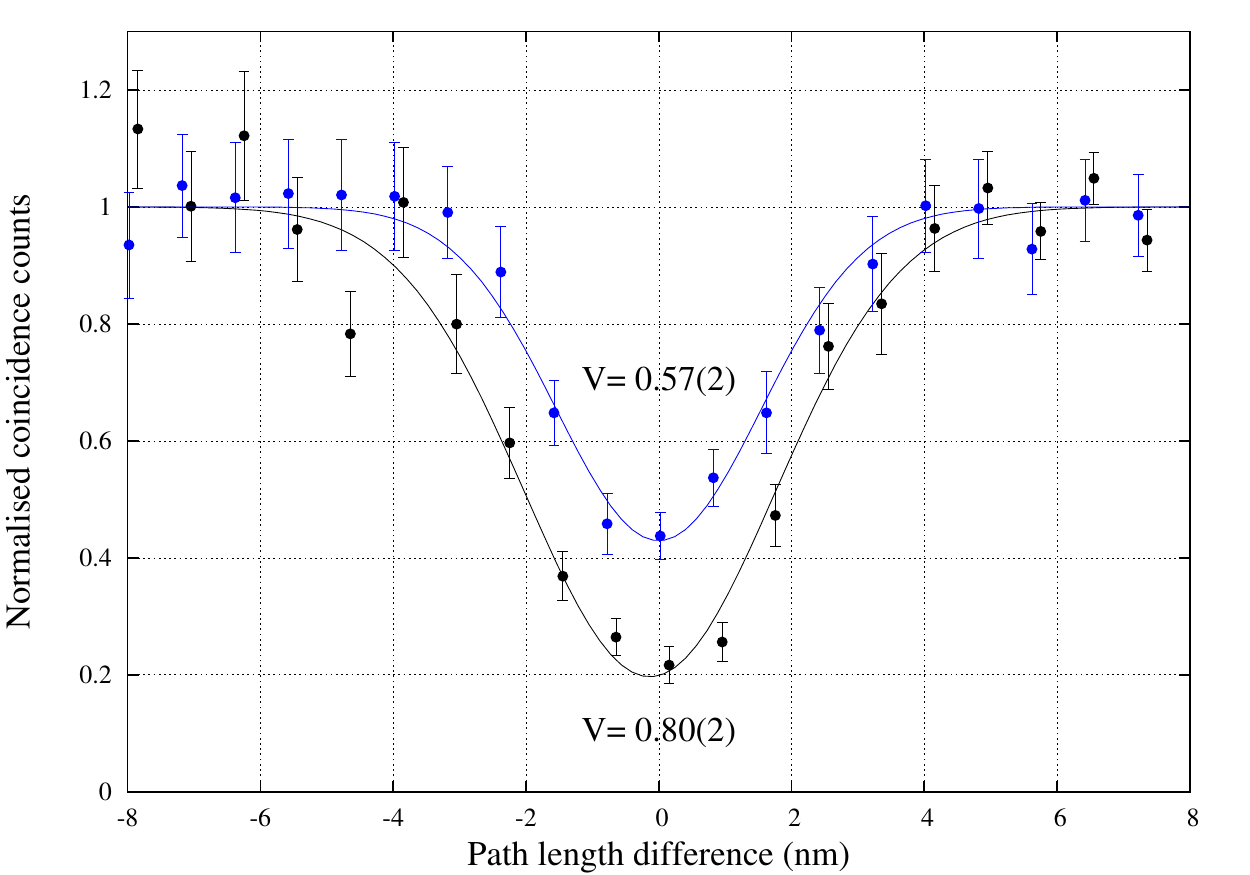}
\caption{HOM dip obtained for two different values of the probability of photon pairs emission for the HSPS. The blue and black data points correspond to $\rm P_1 = $ 0.05 and 0.01, respectively.  The mean number of photons per pulse in the coherent state $\vert \alpha \vert^2 $ is fixed at 0.05. The curves represent the theoretical prediction for two photons with a bandwidth of 80\,MHz and a maximum visibility given by the equation (\ref{eq_maxvis}) for $\rm R=1$.\
\label{fig_dips}}
\end{figure}

The \figurename{~\ref{fig_dips}} shows the experimental results and the theoretical prediction for mean number of photons $|\alpha|^2 = 0.05$ and $\rm P_1 = 0.05$ and 0.01. The theoretical prediction fits perfectly with the experimental measurement for R close to 1, which proves the indiscernibility between the coherent state and the HSP.

\section{Conclusion}

We have demonstrated how difference-frequency generation and spontaneous parametric down conversion in twin crystals can be exploited to generate states of light with common spectral and temporal properties. This was confirmed by performing a Hong, Ou and Mandel like experiment, that showed that the two states overlap perfectly. We utilized sufficiently nonlocal narrowband filtering (26\,GHz in our case) to produce heralded single photons with the same bandwidth of the pump (80\,GHz) in a pure spectral mode. As the CW laser, which seeds the DFG, is monochromatic, the coherent state is constrained to have the same spectral bandwidth as the HSPs but with the advantage that one can work in the non-degenerate wavelength regime. We believe that this technique will be beneficial in a wide range of experiments where one needs to combine single photons  and coherent states.

\section*{Acknowledgment}
We thank B. Sanguinetti, P. Sekatski and H. Zbinden for stimulating discussions. This work was supported in part by the EU project SIQS and the Swiss SNSF project - CR23I2 127118.

\section*{References}
\bibliographystyle{unsrt}
\bibliography{Bib}
\end{document}